# Networks of Reader and Country Status:

# An Analysis of Mendeley Reader Statistics

Robin Haunschild[*], Lutz Bornmann[**], & Loet Leydesdorff[***]

[*] Corresponding author;

Max Planck Institute for Solid State Research,

Heisenbergstr. 1,

70569 Stuttgart, Germany.

Email: R.Haunschild@fkf.mpg.de

[**] Division for Science and Innovation Studies,

Administrative Headquarters of the Max Planck Society,

Hofgartenstr. 8,

80539 Munich, Germany.

Email: bornmann@gv.mpg.de

*** Amsterdam School of Communication Research (ASCoR)

University of Amsterdam,

PO Box 15793,

1001 NG Amsterdam, The Netherlands.

Email: loet@leydesdorff.net

**Abstract**

The number of papers published in journals indexed by the Web of Science core collection is steadily increasing. In recent years, nearly two million new papers were published each year; somewhat more than one million papers when primary research papers are considered only (articles and reviews are the document types where primary research is usually reported or reviewed). However, who reads these papers? More precisely, which groups of researchers from which (self-assigned) scientific disciplines and countries are reading these papers? Is it possible to visualize readership patterns for certain countries, scientific disciplines, or academic status groups? One popular method to answer these questions is a network analysis.

In this study, we analyze Mendeley readership data of a set of 1,133,224 articles and 64,960 reviews with publication year 2012 to generate three different kinds of networks: (1) The network based on disciplinary affiliations of Mendeley readers contains four groups: (i) biology, (ii) social science and humanities (including relevant computer science), (iii) bio-medical sciences, and (iv) natural science and engineering. In all four groups, the category with the addition "miscellaneous" prevails. (2) The network of co-readers in terms of professional status shows that a common interest in papers is mainly shared among PhD students, Master's students, and postdocs. (3) The country network focusses on global readership patterns: a group of 53 nations is identified as core to the scientific enterprise, including Russia and China as well as two thirds of the OECD (Organisation for Economic Co-operation and Development) countries.

**Key words**

Mendeley, network, Pajek, VOSviewer, bibliometrics, altmetrics

# Introduction

Bibliometrics is not only a mature research field, which develops advanced indicators for research evaluation purposes, but also a research field, which studies patterns in science. The best method for studying these patterns is bibliometric networking or science mapping. Here, bibliometric data are used to generate networks of citation relations (e.g. between scholarly journals), networks of co-authorships (e.g. between highly-cited researchers in information science), or networks of co-occurrence relations between keywords, words in abstracts and/ or words in titles (e.g. co-occurrence relations between words in abstracts of papers published in information science) (van Eck & Waltman, 2014). Powerful computers have led to the analysis of large networks, which may include the whole Web of Science (WoS) database from Thomson Reuters (Milojević, 2014). Today, these networks are not only of interest for specialists in bibliometrics or networking, but also for stakeholders from publishers, research institutions, and funding agencies. According to Martin, Nightingale, and Rafols (2014) “network and science-mapping visualizations have considerably enhanced the capacity to convey complex information to users. These tools are now sufficiently mature to be used not only available in academia but also in consultancy and funding organisations” (p. 4). Overviews of publications dealing with networking and mapping have been published, for example, by Börner, Sanyal, and Vespignani (2007), Leydesdorff (2014), and Mingers & Leydesdorff (2015).

In recent years, altmetrics has developed to a popular research field in bibliometrics (Bornmann, 2014). Altmetrics counts and analyzes views, downloads, clicks, notes, saves, tweets, shares, likes, recommends, tags, posts, trackbacks, discussions, bookmarks, and comments to scholarly papers. Altmetrics data reflect different kinds of research impact which has been demonstrated, for example, in the case of Mendeley readership data for social sciences and humanities (Mohammadi & Thelwall, 2013; Mohammadi & Thelwall, 2014; Sud & Thelwall, 2015, in press). Mendeley readership data are essentially bookmarking data. For the sake of simplicity, we refer to the Mendeley data mainly as reader counts. Because it is not clear, what altmetrics counts really measure, most of the studies in this field have calculated the correlation between altmetric counts and citation counts (Bornmann, 2015). A substantial positive correlation points to a certain, but otherwise undefined meaning of altmetrics in a scientific context.

Similar to bibliometric data, altmetric data can not only be used for research evaluation purposes, but also for networking or science mapping. Kraker, Schlögl, Jack, and Lindstaedt (2014) presented a methodology and prototype for creating knowledge domain visualizations based on readership statistics (from Mendeley). Haunschild and Bornmann (2015) generated a readership network which is based on Mendeley readers per (sub-)discipline for a large dataset of biomedical papers.

In this study, we investigate Mendeley readership data for all articles and reviews in WoS where a DOI (digital object identifier) was available from 2012 with the following research questions:

1) Are there differences and similarities between disciplines in bookmarking papers?
2) How do researchers in different career stages differ in terms of bookmarking papers? Which groups of researchers read similar or different papers?
3) Researchers from which countries read papers? Are there patterns of similar readership between specific countries?

We address these questions by studying the network nature of the Mendeley readership data. For this purpose, we generate three different kinds of networks: (1) the network of disciplinary affiliations can show similarities of and differences in the readerships of papers. (2) The status group network shows which status groups (e.g. students, lecturers, or professors) commonly read papers (or not). (3) The country network focuses on global readership patterns: similar and different readings of papers are visualized at the country level.

# Methods

**Dataset used**

Between the 11th and 23rd of December 2014, Mendeley readership statistics for $n_A$ = 1,133,224 articles and $n_R$ = 64,960 reviews were requested via the Application Programming Interface (API), which was made available in 2014, using HTTP GET requests from R (http://www.r-project.org/). An example of the R script is available at http://dx.doi.org/10.6084/m9.figshare.1335688. All papers studied here were published in 2012. The publication year is a compromise of taking a rather recent publication year because Mendeley was founded in 2009 and allowing enough time after publication for reader counts to aggregate. However, as Mendeley reader counts are known to accumulate much faster than citation counts (Maflahi and Thelwall, 2015, in press), we feel justified using the publication year 2012.

The DOIs of the papers in the samples were obtained from the in-house database of the Max Planck Society (MPG) based on the WoS and administered by the Max Planck Digital Library (MPDL). The DOI was used to identify the papers in the Mendeley API. The Mendeley reader counts of 1,074,407 articles (94.8%) and 62,771 reviews (96.6%) were retrieved via the Mendeley API. These percentages are higher than those reported in other studies (Haustein & Larivière, 2014b). The papers which were matched via their DOI in the Mendeley API ($n$ = 1,137,178) are analyzed in the remainder of this study. In total, we recorded 9,352,424 reader counts for articles and 1,335,764 reader counts for reviews.

It is optional for the users of Mendeley to provide their disciplinary affiliations (selecting from predefined sub-disciplines) and location. However, Mendeley does not provide the possible values of country names[1] in the API. Therefore, we used the ISO (International Organization for Standardization) names (see http://countrycode.org) as possible values. Out of the 237 countries we could not find any contributions from 59 countries. However, we are not able to distinguish between a country value which is not possible and a paper with no readers from this country. For example, one is less surprised to find no reader counts for countries like Holy See (Vatican City) than for Singapore.

We retrieved 1,572,240 reader counts (16.8%) for articles and 212,693 reader counts (15.9%) for reviews where the users shared their location information. Country-specific readership information was available for 558,221 (49.3%) articles and 42,935 (66.1%) reviews. The academic status seems to be a mandatory piece of information, as the total number of Mendeley readers found agrees with the status-specific readership information. The self-assigned sub-discipline is not mandatory but most Mendeley users provide it in our sample set. Only 4,924

[1] The country names in the Mendeley web frontend are standardized. The user provides the city and Mendeley proposes different city-country combinations from which the user can choose.

(0.05%) of the Mendeley article readers and 531 (0.04%) review readers did not share their (sub-) disciplinary affiliation.

**Software and Statistics**

The data was organized at three levels of aggregation:

a) groups of individual readers who bookmark the papers, in terms of disciplinary affiliations;
b) groups of readers in terms of their professional status (Professor, PhD student, postdoc, etc.);
c) groups of readers in terms of their countries as provided by Mendeley readers in their profile.

The Mendeley bookmarking can be considered as referencing, and then the analysis of this Mendeley data is analogous to bibliographic coupling in bibliometrics (Kessler, 1963). Although being analogous to bibliographic coupling, the bookmark coupling provides different kinds of information in comparison to bibliographic coupling: First, bibliographic coupling is based on the references in the paper, while Mendeley reader counts are similar to times cited data and thus reflect the citing-side perspective. Second, bibliographic coupling captures only authors of papers which are indexed in a citation index. There is no necessary relationship between authoring and reading papers: some people read more literature and author few papers or write more monographs.

Bookmark couplings also capture users of Mendeley who author fewer papers or publish in journals which are not indexed in popular citation indices. However, bookmark coupling has another bias, as not everyone uses Mendeley to bookmark papers. Both methods (bibliographic and bookmark coupling) are interesting to analyze networks of publications. They complement each other.

In each of the three analyses, the largest component is extracted, and further analyzed using the community finding algorithm of Blondel, Guillaume, Lambiotte, and Lefebvre (2008). Pajek is used for the network analysis. Default values were used during construction and analysis of all networks. All reader counts are weighted equally (Pajek option "unweighted"), and each network connection is counted as a single co-bookmarking event. The results are visualized using VOSviewer.

# Results

*a. statistical parameters*

The three networks presented below are compared in terms of network statistics in Table 1.

**Table 1**: Statistics of the full networks of disciplinary affiliations, countries, and status groups

| Statistical parameter | Disciplinary affiliation | Country | Status group |
|---|---|---|---|
| Number of vertices | 465 | 178 | 13 |
| Average degree | 243.45 | 76.02 | 13.00 |
| Degree centralization | 0.47 | 0.54 | 0.00 |
| Density | 0.53 | 0.43 | 1.00 |
| Closure | 0.72 | 0.70 | 1.00 |
| Average distance | 1.48 | 1.58 | 1.00 |
| Standard deviation of average distance | 0.50 | 0.51 | 0.00 |
| Diameter | 3 | 3 | 1 |
| Compactness | 0.76 | 0.71 | 1.00 |
| Modularity | 0.25 | 0.02 | 0.00 |

The network among the 13 status groups is fully connected; but we will discuss the relative weights of the relations in the following. The other two networks are very different in nature, despite the seeming similarity in some of these parameters.

*b. disciplinary affiliations*

Among the disciplinary affiliations, 470 could be distinguished in this data, of which 465 (98.94%) form a largest component. The five affiliations which were not connected are: “Judaism”, “Catholicism”, “Transport Law”, “Entertainment, Sports and Gaming Law”, and “Air and Space Law”. These five affiliations belong to the humanities (theology and law, respectively). We found a total of three reader counts for “Judaism” and one reader count each for the other four disconnected disciplinary affiliations. Only very few researchers in these disciplines seem to use Mendeley. Similar results have been reported by Jeng et al. (2015) who reported that they “did not see many group users from the humanities and other related fields” (p. 898).

The 465 affiliations in the main component can be sorted into four groupings by the community-finding algorithm of Blondel *et al*. (2008); the modularity—a measure for the quality of the clustering between zero and one—is $Q = 0.25$ (c. f. Table 1). The four groups are, respectively:

1. 54 affiliations in biology and the geo-sciences;
2. 256 affiliations mainly in the social sciences and the humanities (Figure 1);
3. 71 affiliations in the bio-medical sciences (Figure 2);
4. 84 affiliations in the natural sciences and engineering (also included in Figure 2).

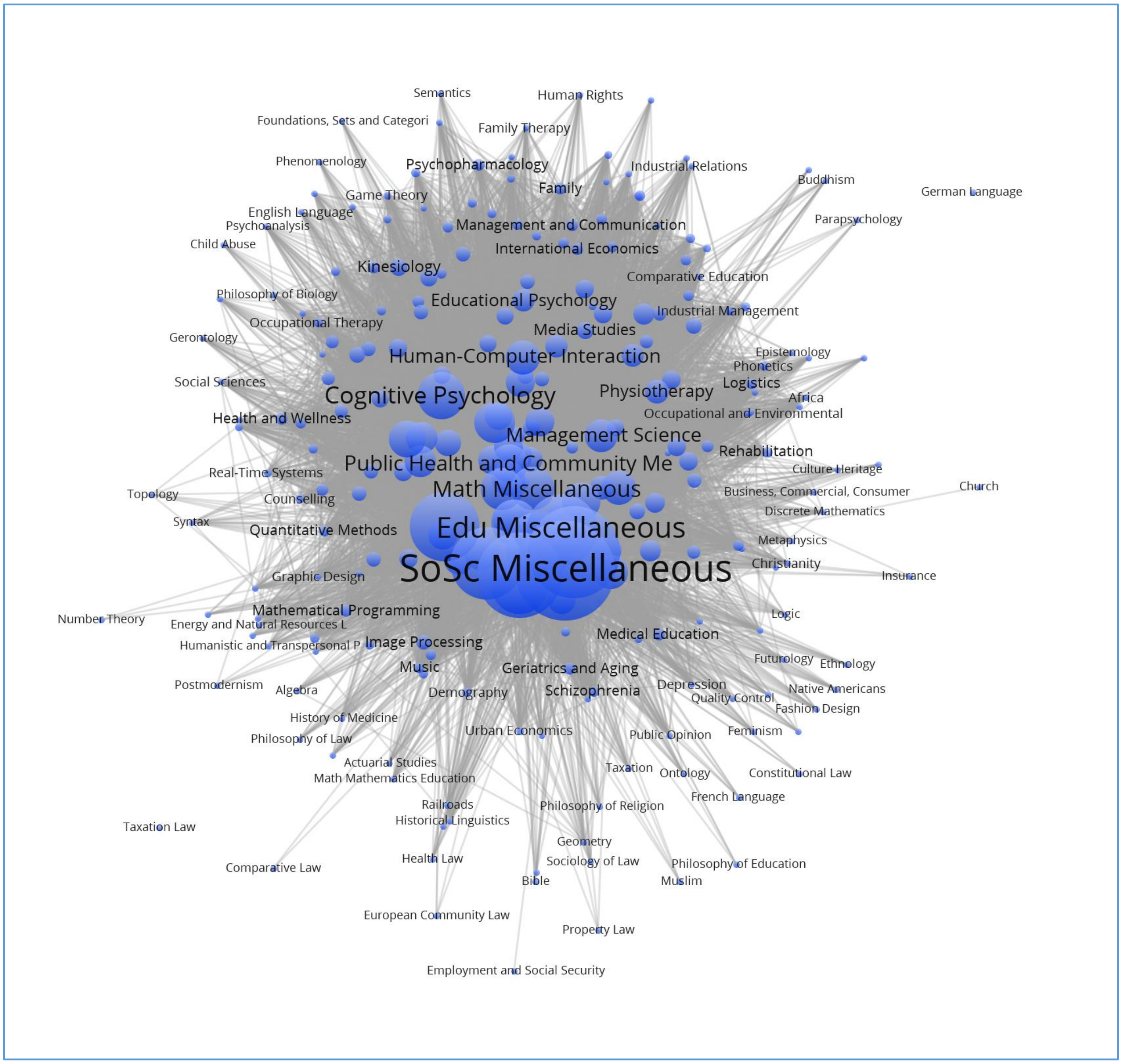

**Figure 1**. 256 affiliations, mainly in the social sciences and the humanities (Group 2). This figure can be web-started at
http://www.vosviewer.com/vosviewer.php?map=http://www.leydesdorff.net/mendeley/fig1_map.txt&network=http://www.leydesdorff.net/mendeley/fig1_net.txt&n_lines=10000

Figure 1 shows 256 sub-discipline affiliations of Mendeley readers (group 2) with their connections in the social sciences and humanities. Zahedi & van Eck (2014) have found similar results. They reported that Mendeley users are most active in the biomedical sciences, life sciences, and social sciences. The network shown in Figure 1 also includes some reading in the computer sciences and mathematics. The relation seems to be via cognitive psychology, artificial intelligence, etc. The humanities are positioned more at the periphery of this set. The sub-disciplines “Taxation law” and “German language” are not directly connected to this sub-group, but nevertheless sorted into it by the community-finding algorithm. The number of readers providing bookmarks to these sub-disciplines is low.

Figure 2 shows 71 sub-discipline affiliations in the bio-medical sciences (group 3) and 84 sub-discipline affiliations in the natural sciences and engineering (group 4). We do not show the links in order to keep the distinction between the two sets of nodes (with different colors) focal to the visualization. A version with the network links visible can be web-started from http://www.vosviewer.com/vosviewer.php?map=http://www.leydesdorff.net/mendeley/fig2_map.txt&network=http://www.leydesdorff.net/mendeley/fig2_net.txt&n_lines=10000

It is somewhat surprising to see the sub-disciplines “regional law” and “Latin” sorted into the network of mainly bio-medical sciences in Figure 2. As the links in the web-started version show, these bookmarks have many links to several sub-disciplines within the bio-medical network.

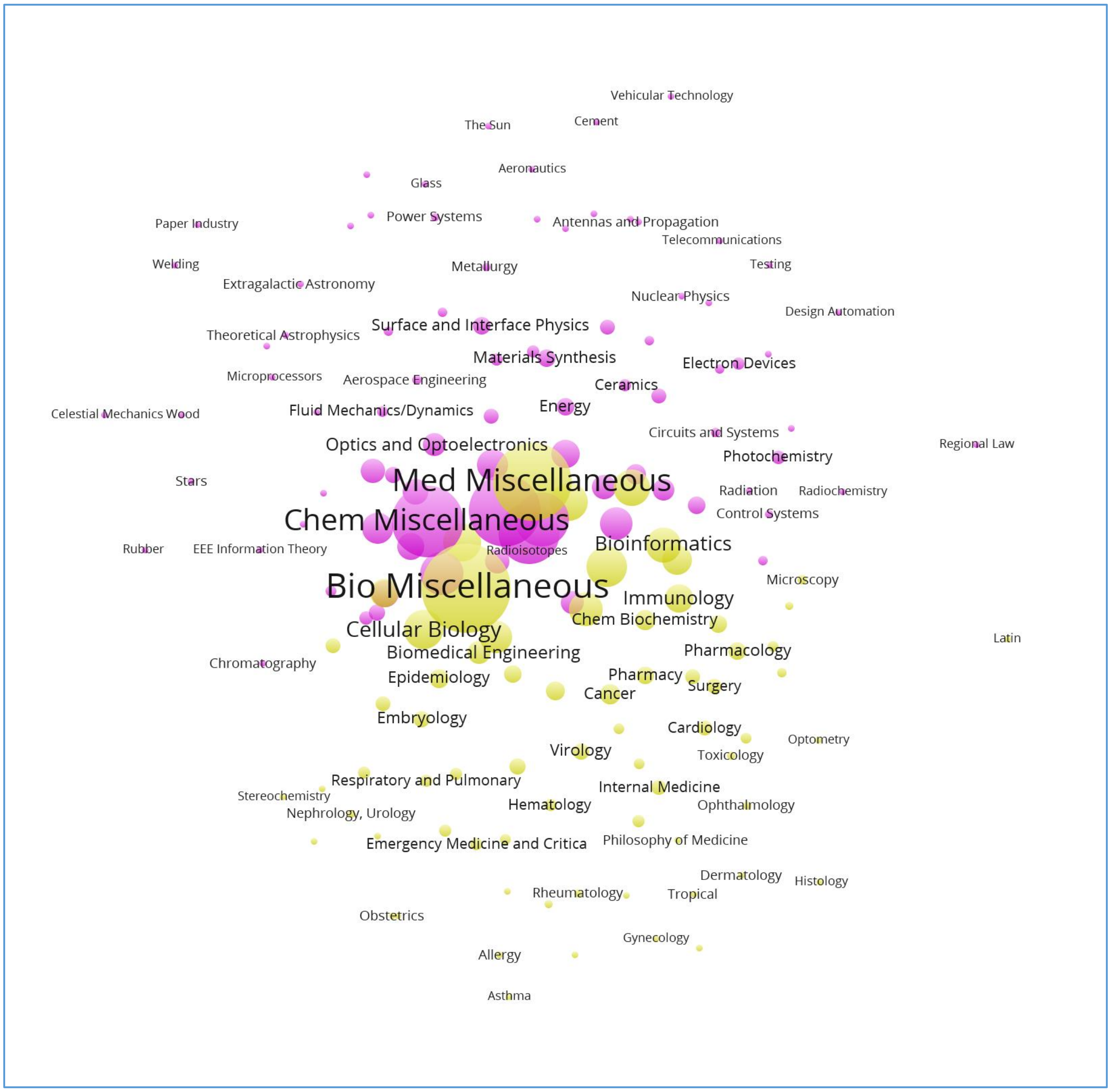

**Figure 2**: 71 affiliations in the bio-medical sciences (yellow) and 84 affiliations in the natural sciences and engineering (pink).

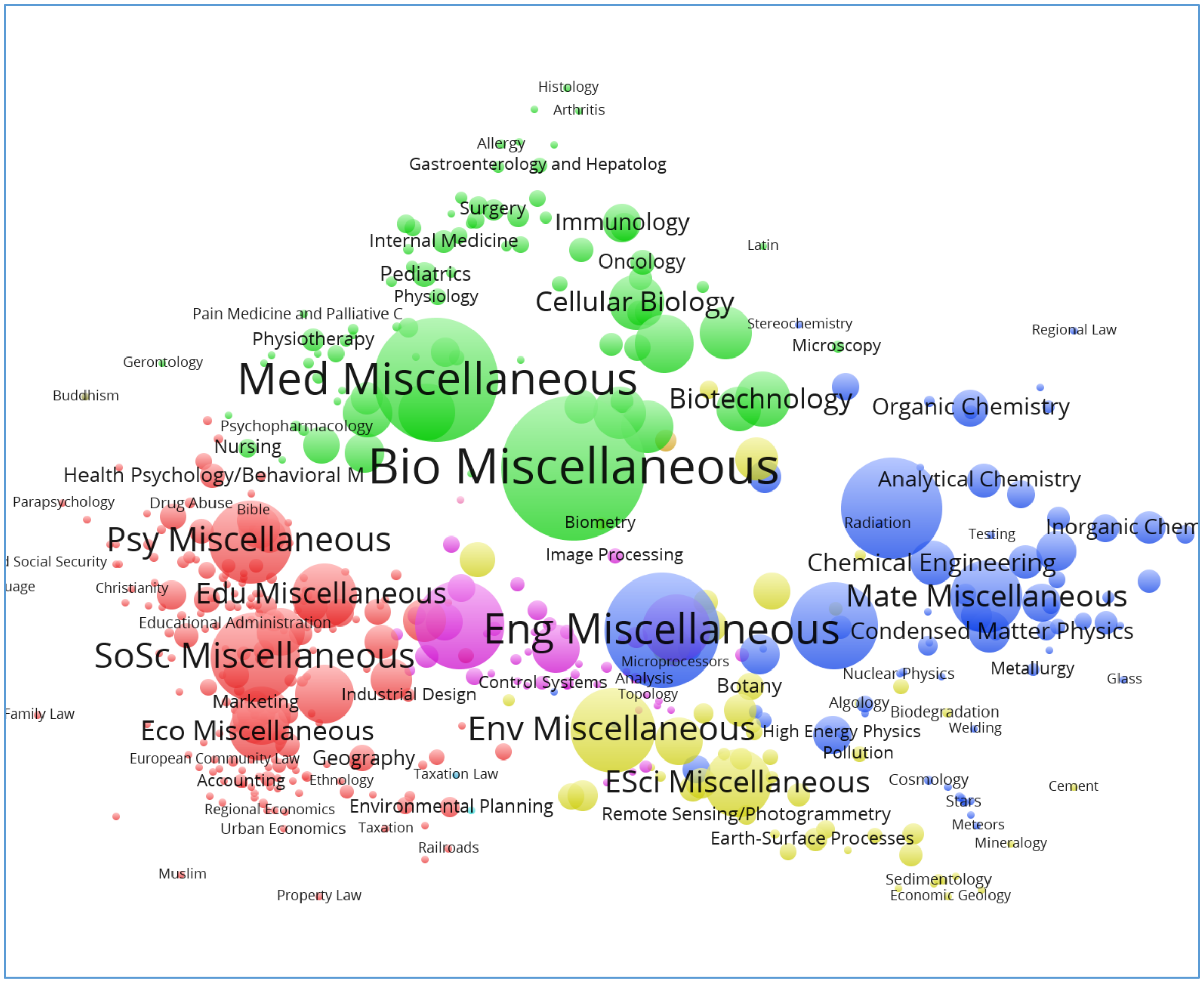

**Figure 3**: Four communities (colors) of affiliations among co-bookmarking readers ($Q$ = 0.25).

Figure 3 visualizes the entire network of sub-disciplinary affiliations. It shows that the core set is occupied by readers who characterize themselves as “miscellaneous” readers from different disciplines such as “biology miscellaneous”, “environmental science miscellaneous”, etc. The social sciences (“miscellaneous”) are one among these reading communities. The humanities, however, are placed more in the periphery. The algorithmically generated distinctions among the four groups (using Blondel *et al.*, 2008) cannot be clearly distinguished using this projection, because the domains overlap when projected on a two-dimensional plane. The figure is therefore based on the mapping of VOSviewer in this case. This figure can be web-started from

http://www.vosviewer.com/vosviewer.php?map=http://www.leydesdorff.net/mendeley/fig3_map.txt&network=http://www.leydesdorff.net/mendeley/fig3_net.txt&n_lines=1000

*c. status hierarchy*

Mendeley users have to assign one of 14 predefined status groups to themselves. Some of these status groups seem redundant, such as “Student PhD”, “Student Post-Graduate”, and “Doctoral Student”. Any merging or regrouping of these status groups, however, would be a rather arbitrary choice (Haustein & Larivière, 2014a). We analyze the Mendeley reader counts in the status groups as provided by Mendeley and discuss these issues in the light of the results.

Figure 4 shows that a common interest in papers is mainly shared among PhD students, Master’s students, and postdocs. Other studies confirm the dominant position of these groups (Zahedi, Costas, & Wouters, 2014). In this study, researchers at academic institutions follow, but less so when compared with researchers at non-academic institutions. Lecturers and Senior Lecturers are less involved than Professors. Librarians hardly participate in this network. Note that this network is not modularized at all ($Q$ = 0.0, c. f. Table 1). All groups are fully connected to all (13) other groups.

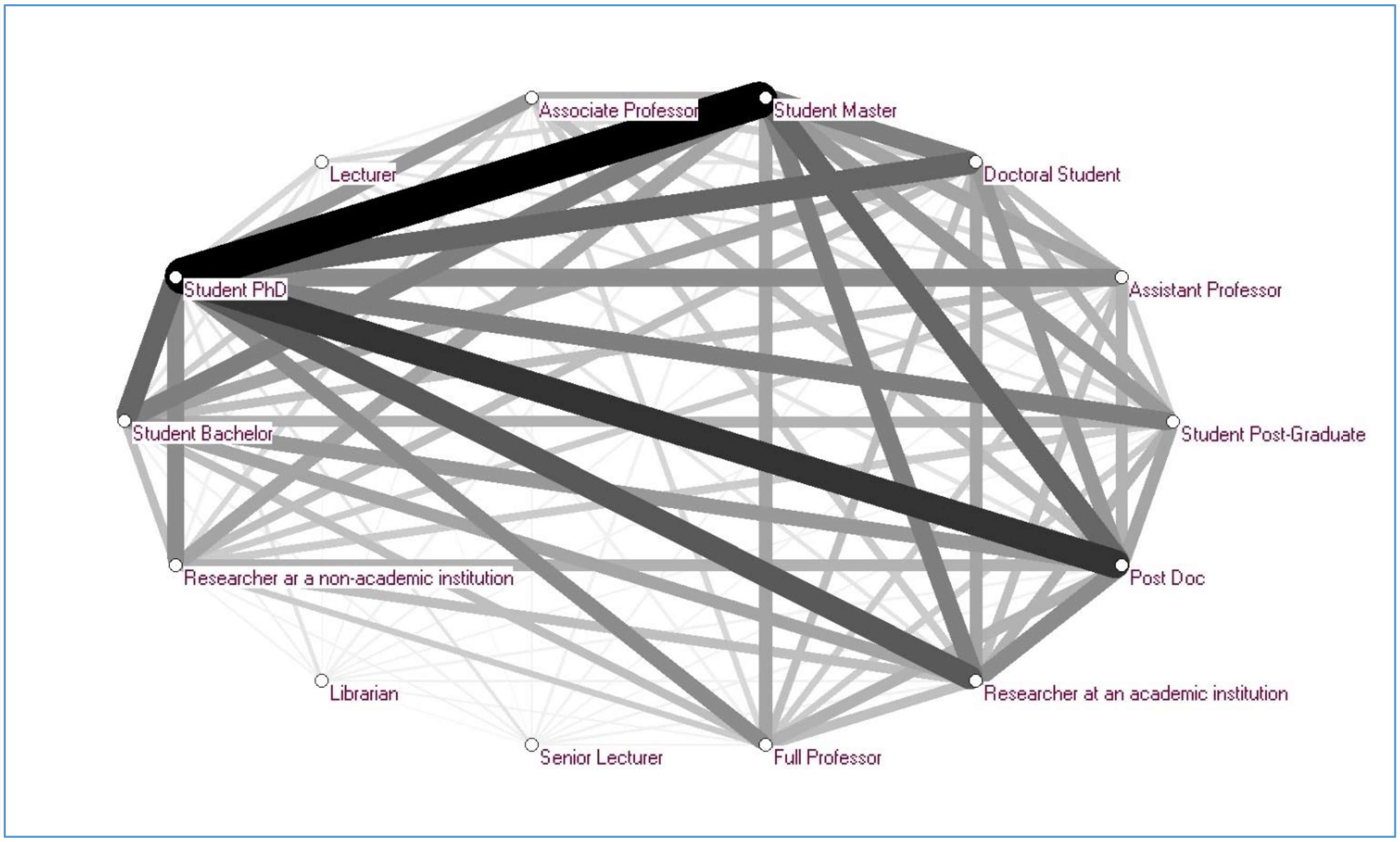

**Figure 4**: Network of co-readers in terms of professional status.

**Table 2**: Eigenvector centralities and absolute number of reader counts *N* of different status groups among networked Mendeley users (using the Hubs & Authorities routine in Pajek).

| Status group | Eigenvector centrality | *N* |
|---|---|---|
| Student PhD | 0.45 | 3,364,144 |
| Student Master | 0.39 | 1,514,606 |
| Post Doc | 0.34 | 1,148,860 |
| Researcher at an academic institution | 0.30 | 667,995 |
| Doctoral Student | 0.29 | 616,738 |
| Student Bachelor | 0.28 | 678,839 |
| Student Post-Graduate | 0.25 | 482,784 |
| Assistant Professor | 0.23 | 409,591 |
| Researcher at a non-academic institution | 0.23 | 444,874 |
| Full Professor | 0.22 | 378,685 |
| Associate Professor | 0.20 | 316,606 |
| Lecturer | 0.10 | 126,848 |
| Librarian | 0.06 | 77,046 |
| Senior Lecturer | 0.05 | 63,345 |

Table 2 shows eigenvector centralities of the different status groups among networked Mendeley users (Bonacich, 1972). Groups with high eigenvector centrality (in this case, students) are more central, because they share their interests in publications with many other groups, while recursively taking into account the (eigenvector) centrality of these other groups (de Nooy, Mrvar, and Batagelj, 2011). However, the very high eigenvector centrality is probably to some extend due to the fact that students (especially PhD and Master) and postdocs form by far the largest status groups. This is in agreement with previous studies which also found that students and postdocs represent the largest user status groups at Mendeley (Bornmann & Haunschild, 2015; Mohammadi, Thelwall, Haustein, & Larivière, 2015). Senior Lecturers – a group with the lowest eigenvector centrality– seem to be interested in publications different from the other status groups. However, the eigenvector centrality is strongly influenced by the absolute number of reader counts. The Spearman rank correlation coefficient between eigenvector centrality and reader counts is 0.986.

Note that the status indication may be different among nations. For example, the ranks of "Assistant Professor" and "Lecturer" are virtually non-existent in some countries. On the other hand side, ranks such as "Reader" (sometimes different from "Lecturer") and "Habilitand"[2] are not covered by the Mendeley classification system. The data suggests that Mendeley readers in the career stages "Reader" and "Habilitand" assign the status "Assistant Professor" to themselves, as this is the highest populated among the professorship categories. Furthermore, some status groups seem redundant, e.g. "doctoral student", "Student Post-Graduate", and

[2] "Habilitand" is a status in German-speaking countries for those working on a "Habilitation" as a second PhD which provides teaching rights in the university.

"Student PhD". However, most Mendeley readers who are working on a doctoral thesis identify themselves as "Student PhD".

*d. decomposition in terms of nations*

Among the 200+ countries in the world, 178 countries are indicated among the readership of Mendeley that actively bookmarked records in this database. These countries are all connected with an average degree of 76.02 which means that on average each node is linked to 76 (42.7%) other nodes in the network of 178 nodes; the density of the network is 0.43 (c. f. Table 1). The eigenvector centralities of the countries vary only between 0.055 and 0.077. This small variation of eigenvector centrality between countries is probably due to the high connectivity of the countries although there is a large variation of reader counts from 1 (Liberia) to 396,198 (USA).

The community-finding algorithm distinguishes four groups. However, the modularity among these four groups is low (Q = 0.02, c. f. Table 1) because of cross-group network connections:

1. A group of 53 nations that are core to the scientific enterprise, including Russia and China as well as two thirds of the OECD countries (Figure 5). The OECD member states Chile, Greece, Iceland, Mexico, New Zealand, Norway, Portugal, Slovak Republic, Slovenia, and Turkey are not part of this group. They are part of the second group.
2. A largest group of 115 nations centered around Brazil and India, but including also Norway (Figure 6).
3. A group of ten small nations with "Niger" and Nigeria" as the central core.
4. The smallest group with only "Guinea" and "Guinea Bissau".

Figures 5 and 6 show the country groups 1 and 2. A version of Figure 6 can be web-started at http://www.vosviewer.com/vosviewer.php?map=http://www.leydesdorff.net/mendeley/fig6_map.txt&network=http://www.leydesdorff.net/mendeley/fig6_net.txt&n_lines=10000&label_size=1.0&label_size_variation=0.34. As in the case of Figure 5, one can run mapping and clustering of the subsets in VOSviewer for obtaining more details. The results of these finer-grained decompositions do not obviously make sense to us.

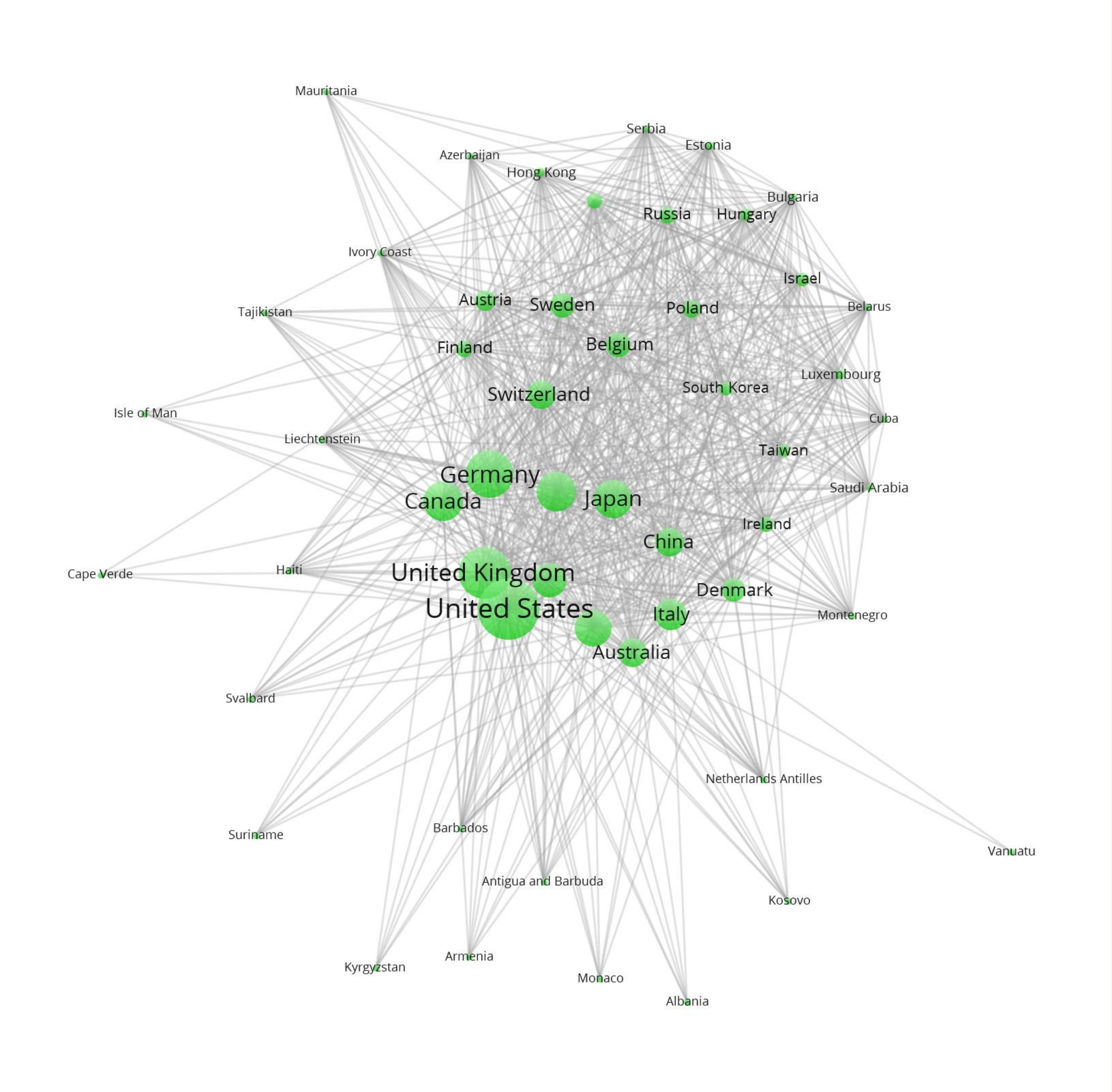

**Figure 5**: Group of 53 nations. The unlabeled circles next to the UK and the US indicate the Netherlands and Spain, respectively. The unlabeled circle between Russia and Hong Kong is the Czech Republic. A version with the labels visible can be web-started from [http://www.vosviewer.com/vosviewer.php?map=http://www.leydesdorff.net/mendeley/fig5_map.txt&network=http://www.leydesdorff.net/mendeley/fig5_net.txt&n_lines=10000&label_size=1.0&label_size_variation=0.34](http://www.vosviewer.com/vosviewer.php?map=http://www.leydesdorff.net/mendeley/fig5_map.txt&network=http://www.leydesdorff.net/mendeley/fig5_net.txt&n_lines=10000&label_size=1.0&label_size_variation=0.34).

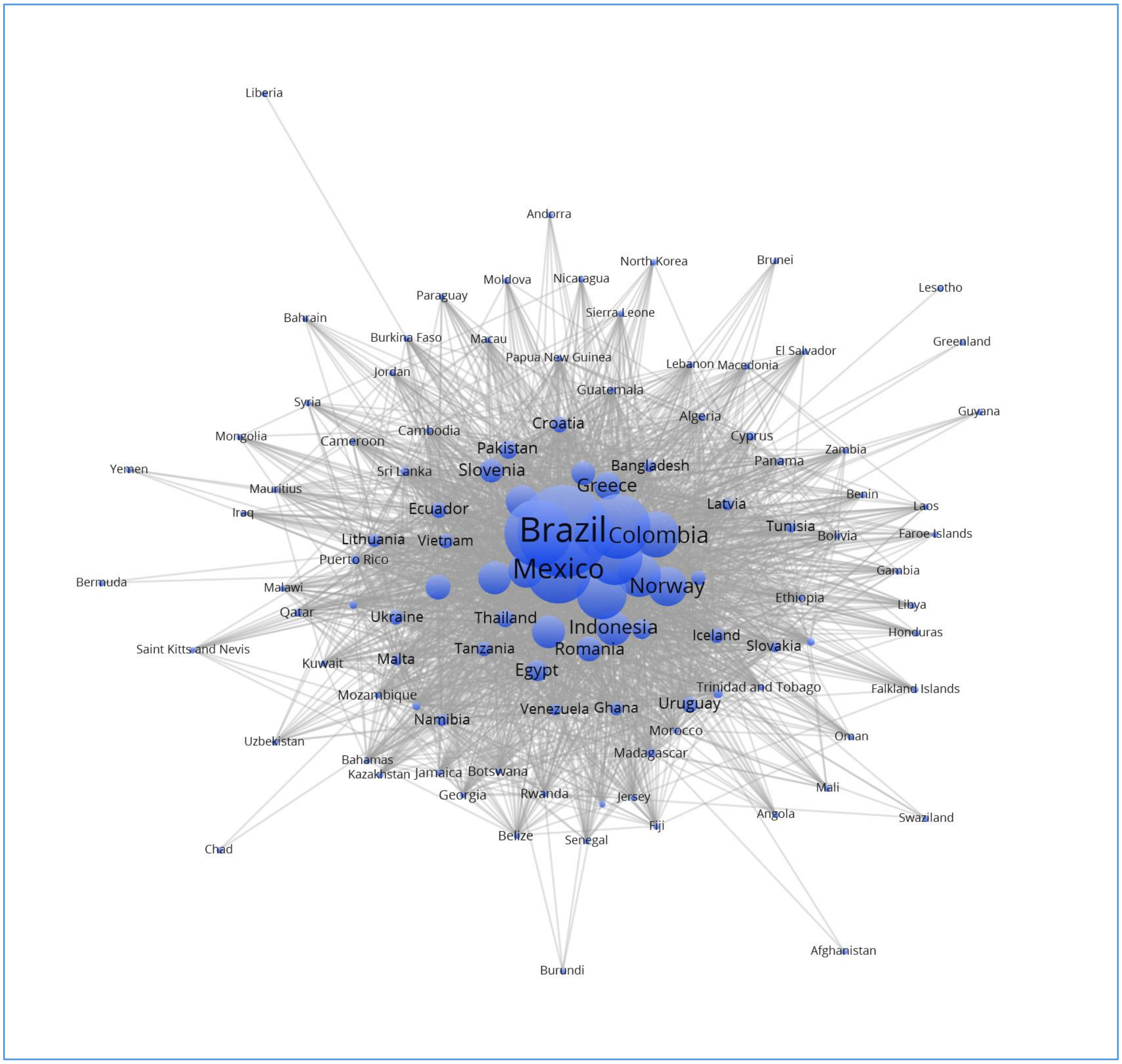

**Figure 6**: 115 countries in the second group of nations.

# Discussion

Networks are one of the most important and popular methods to analyze bibliometric data. In this study, we explored whether Mendeley data can also be successfully used as a data source for network analysis. Only a few attempts have been made up to now to analyze the rich Mendeley data using network analyses techniques. It is a great advantage that this data can be retrieved for comprehensive publication sets using an API. Thus, one can download readership data on a large scale; this data is very suitable for network analyses. We encourage other researchers to use Mendeley data for larger publication sets in order to inspect usage structure of publications (Gunn, 2014).

The Mendeley readership networks were generated by using different types of user information: their (1) discipline, (2) status, and (3) country. All three information sources could be used to produce meaningful network results. In terms of disciplines, first, we found four groups: (1) biology, (2) social science and humanities (including relevant computer science), (3) bio-medical sciences, and (4) natural science and engineering. In all four groups, the category with the addition "miscellaneous" prevails. Probably, the readers who identify themselves with cross-disciplinary research interests are more inclined to generate these "bookmark couplings" than more specifically specialized readers. The pronounced position of the social sciences and the humanities was not expected. Some sub-disciplines, e.g. "Judaism" and "Catholicism", are disconnected from the other sub-disciplines.

The decomposition in terms of status hierarchies within the network makes clear that this hierarchy is inversed in Mendeley. The lead among the users is taken by students working on theses. More than professionals, students have time to explore the literature beyond their specialization. Lecturers and Senior Lecturers entertain a different reading pattern, given their primary tasks in education. Librarians make use of Mendeley (and scholarly literature) differently from researchers. Students also have the highest eigenvector centrality in the network which indicates that they have a strong bookmark coupling when compared with other status groups. However, the reader count distribution is skewed. The calculated eigenvector centralities correlate strongly with the absolute number of observed reader counts.

The decomposition in terms of nations highlights the worldwide divide between developed and less-developed nations. A similar prevailing divide was recently also found in portfolio analysis of journal literature by Leydesdorff, Heimeriks, & Rotolo (in press). More fine-grained delineations can partially be recognized as regional, but could not always be provided with an obvious interpretation.

The academic status information is provided by every Mendeley user and nearly every Mendeley users provides (sub-) discipline information, while only a minority of Mendeley users seems to provide their location. This makes it more difficult to analyze the reader counts broken down by countries. Some Mendeley academic status groups seem redundant, while others seem to be tailored to the US system. Surprisingly, the vast majority of Mendeley readers assign the miscellaneous sub-discipline of their main discipline to themselves. It is not clear to which extent Mendeley users assign the precise sub-discipline, status, and location information to themselves and whether they update this information regularly. Despite these shortcomings of the Mendeley classification system and the quality of information the users provide, the network analyses of Mendeley reader counts from three different perspectives produced interesting insights in readership patterns. This shows that useful network analysis can be performed using Mendeley readership counts.

## Conclusions

In this study, we analyzed Mendeley readership data of a set of 1,074,407 articles and 62,771 reviews with publication year 2012 to generate three different kinds of networks: (1) The network based on disciplinary affiliations of Mendeley readers contains four groups: (i) biology, (ii) social science and humanities (including relevant computer science), (iii) bio-medical sciences, and (iv) natural science and engineering. In all four groups, the category with the addition “miscellaneous” prevails. (2) The network of co-readers in terms of professional status shows that Mendeley is mainly shared among PhD students, Master’s students, and postdocs. (3) The country network focusses on global readership patterns: It identifies a group of 53 nations that are core to the scientific enterprise, including two thirds of the OECD countries as well as Russia and China.

## Acknowledgements

The bibliometric data used in this paper are from an in-house database developed and maintained by the Max Planck Digital Library (MPDL, Munich) and derived from the Science Citation Index Expanded (SCI-E), Social Sciences Citation Index (SSCI), Arts and Humanities Citation Index (AHCI) prepared by Thomson Reuters (Philadelphia, Pennsylvania, USA).

# References

Blondel, V. D., Guillaume, J. L., Lambiotte, R., & Lefebvre, E. (2008). Fast unfolding of communities in large networks. *Journal of Statistical Mechanics-Theory and Experiment*. doi: 10.1088/1742-5468/2008/10/P10008.

Börner, K., Sanyal, S., & Vespignani, A. (2007). Network science. *Annual Review of Information Science and Technology, 41*(1), 537-607. doi: 10.1002/aris.2007.1440410119.

Bornmann, L. (2014). Do altmetrics point to the broader impact of research? An overview of benefits and disadvantages of altmetrics. *Journal of Informetrics, 8*(4), 895-903. doi: 10.1016/j.joi.2014.09.005.

Bornmann, L. (2015). Alternative metrics in scientometrics: A meta-analysis of research into three altmetrics. *Scientometrics, 103*(3), 1123-1144.

Bornmann, L., & Haunschild, R. (2015). Which people use which scientific papers? An evaluation of data from F1000 and Mendeley. *Journal of Informetrics, 9*(3), 477-487. doi: 10.1016/j.joi.2015.04.001.

Bonacich, P. (1972). Factoring and weighting approaches to status scores and clique identification. *Journal of Mathematical Sociology,* 2(1), 113-120.

de Nooy, W., Mrvar, A., & Batagelj, V. (2011). *Exploratory social network analysis with Pajek*. New York, NY, USA: Cambridge University Press.

Gunn, W. (2014). Mendeley: Enabling and understanding scientific collaboration. *Information Services and Use, 34*(1), 99-102. doi: 10.3233/ISU-140738.

Haustein, S., & Larivière, V. (2014a). *Mendeley as a Source of Readership by Students and Postdocs? Evaluating Article Usage by Academic Status.* Paper presented at the Proceedings of the IATUL Conferences. Paper 2.

Haustein, S., & Larivière, V. (2014b). A multidimensional analysis of Aslib proceedings – using everything but the impact factor. *Aslib Journal of Information Management, 66*(4), 358-380. doi: doi:10.1108/AJIM-11-2013-0127.

Haunschild, R., & Bornmann, L. (2015). F1000Prime: an analysis of discipline-specific reader data from Mendeley [v2; ref status: awaiting peer review, http://f1000r.es/50a]. *F1000Research, 4*(41).

Jeng, W., He, D. Q., & Jiang, J. P. (2015). User Participation in an Academic Social Networking Service: A Survey of Open Group Users on Mendeley. *Journal of the Association for Information Science and Technology, 66*(5), 890-904. doi: 10.1002/asi.23225.

Kessler, M. M. (1963). Bibliographic coupling between scientific papers. *American Documentation, 14*(1), 10-25.

Kraker, P., Schlögl, C., Jack, K., & Lindstaedt, S. (2014). Visualization of Co-Readership Patterns from an Online Reference Management System. Retrieved October 17, 2014, from http://arxiv.org/abs/1409.0348

Leydesdorff, L. (2014). Science Visualization and Discursive Knowledge. In B. Cronin & C. Sugimoto (Eds.), *Beyond Bibliometrics: Harnessing Multidimensional Indicators of Scholarly Impact* (pp. 167-185). Cambridge MA: MIT Press.

Leydesdorff, L., Heimeriks, G., & Rotolo, D. (2015, in press). Journal Portfolio Analysis for Countries, Cities, and Organizations: Maps and Comparisons. *Journal of the Association for Information Science and Technology*.

Maflahi, N. and Thelwall, M. (2015, in press). When are readership counts as useful as citation counts? Scopus versus Mendeley for LIS journals. *Journal of the Association for Information Science and Technology*.
Martin, B., Nightingale, P., & Rafols, I. (2014). *Response to the Call for Evidence to the Independent Review of the Role of Metrics in Research Assessment*.
Milojević, S. (2014). Network Analysis and Indicators. In Y. Ding, R. Rousseau & D. Wolfram (Eds.), *Measuring Scholarly Impact* (pp. 57-82): Springer International Publishing.
Mingers, J., & Leydesdorff, L. (2015, early view). A Review of Theory and Practice in Scientometrics. *European Journal of Operational Research*, 1-19. doi: 10.1016/j.ejor.2015.04.002.
Mohammadi, E., & Thelwall, M. (2013). Assessing the Mendeley readership of social science and humanities research. In J. Gorraiz, E. Schiebel, C. Gumpenberger & M. Ho (Eds.), *Proceedings of ISSI 2013 Vienna: 14th International society of scientometrics and informetrics conference* (pp. 200-214). Vienna, Austria: Austrian Institute of Technology GmbH.
Mohammadi, E., & Thelwall, M. (2014). Mendeley readership altmetrics for the social sciences and humanities: Research evaluation and knowledge flows. *Journal of the Association for Information Science and Technology*, 65(8), 1627-1638. doi: 10.1002/asi.23071.
Mohammadi, E.. Thelwall, M., Haustein, S., and Larivière, V. (2015). Who reads research articles? An altmetrics analysis of Mendeley user categories. *Journal of the Association for Information Science and Technology*, 66(9), 1832–1846. doi: 10.1002/asi.23286.
Sud, P. & Thelwall, M. (2015, in press). Not all international collaboration is beneficial: The Mendeley readership and citation impact of biochemical research collaboration. *Journal of the Association for Information Science and Technology*.
van Eck, N., & Waltman, L. (2014). Visualizing Bibliometric Networks. In Y. Ding, R. Rousseau & D. Wolfram (Eds.), *Measuring Scholarly Impact* (pp. 285-320): Springer International Publishing.
Zahedi, Z., Costas, R., & Wouters, P. (2014). Assessing the Impact of Publications Saved by Mendeley Users: Is There Any Different Pattern Among Users? Proceedings of the IATUL Conferences. Paper 4. Retrieved September 10, 2014, from http://docs.lib.purdue.edu/iatul/2014/altmetrics/4
Zahedi, Z., & Eck, N. J. v. (2014). Visualizing readership activity of Mendeley users using VOSviewer. Retrieved July 8, 2014, from http://figshare.com/articles/Visualizing_readership_activity_of_Mendeley_users_using_VOSviewer/1041819